\documentclass[aps,amssymb,amsmath,prl,twocolumn,superscriptaddress]{revtex4-1}

\usepackage{graphicx}
\usepackage{amssymb}
\usepackage{epstopdf}
\usepackage{amsmath}
\usepackage{amsthm}
\usepackage{units}

\def\be{\begin{equation}}   \def\ee{\end{equation}}
\def\eq#1{{Eq.(\ref{#1})}}    \def\fig#1{{Fig.\ref{#1}}}

\def\kT{k_{{}_{\rm B}}T}

\begin{document}

\title{Reversibility of Red blood Cell deformation.}
\author{Maria Zeitz, Pierre Sens\\ \vspace{0.2cm}{\em {\small UMR 7083 ``Gulliver'', CNRS\\ ESPCI, 10 rue Vauquelin, 75231 Paris Cedex 05 - France\\ maria.zeitz@uni-dortmund.de, pierre.sens@espci.fr}}}
\date{\today}

\begin{abstract}
The ability of cells to undergo reversible shape changes is often crucial to their survival. For Red Blood Cells (RBCs), irreversible alteration of the cell shape and flexibility often causes anemia. Here we show theoretically that RBCs may react irreversibly to mechanical perturbations because of tensile stress in their cytoskeleton. The transient polymerization of protein fibers inside the cell seen in sickle cell anemia or a transient external force can  trigger the formation of a cytoskeleton-free membrane protrusion of $\mu$m dimensions. The complex  relaxation kinetics of the cell shape is shown to be responsible for selecting the final state once the perturbation is removed, thereby controlling the reversibility of the deformation. In some case, tubular protrusion are expected to relax via a peculiar ``pearlingÕÕ instability''.
\end{abstract}

\maketitle

Red Blood Cells (RBCs) have been extensively studied by physicists as a  relatively simple example of biological cells  \cite{evans:1994a,lim:2008}. Their mechanical properties reflect the structure of the cell interface, including the (fluid) plasma membrane (PM) and the  cytoskeleton (CSK), 
a two-dimensional  network of flexible spectrin filaments connected to the membrane through node complexes \cite{evans:1994a}. This composite mechanics is responsible for a remarkable variety of equilibrium shape (stomatocyte-discocyte-echinocyte) \cite{lim:2008,sens:2007}. RBCs also exhibit a complex dynamical response to mechanical stress, characterized by visco-elasticity, and plastic deformation at high strain \cite{markle:1983,fischer:2004}.

 Some RBCs undergo major deformation during the respiratory cycle, and it appears important to understand the extent to which such large deformation might be reversible. In sickle-cell anemia, a mutation in the hemoglobin (Hg) gene leads to the formation of Hg fibers inside the cell and results in sickle shaped cells  \cite{noguchi:1981}. The fibers depolymerize in the lung because of oxygen intake, but polymerize again as oxygen is released. The cells loose their flexibility and may obstruct small blood capillaries, with serious medical consequences \cite{serjeant:1997}.  Unpublished observations suggest that in some case, rapid fiber depolymerization produces a large membrane bleb that fails to reincorporate the cell \cite{Note1}. A (very) long membrane tether extracted  from a RBC by a external force ({\em e.g} applied by optical tweezers) may also sometimes fail to retract into the cell when the force is switched off \cite{Note2}. Such events, if they occurred {\em in-vivo}, could lead to repeated loss of cell membrane area and considerably shorten the lifetime of RBCs.
 
 \begin{figure}[b]
	\includegraphics[width=7cm ]{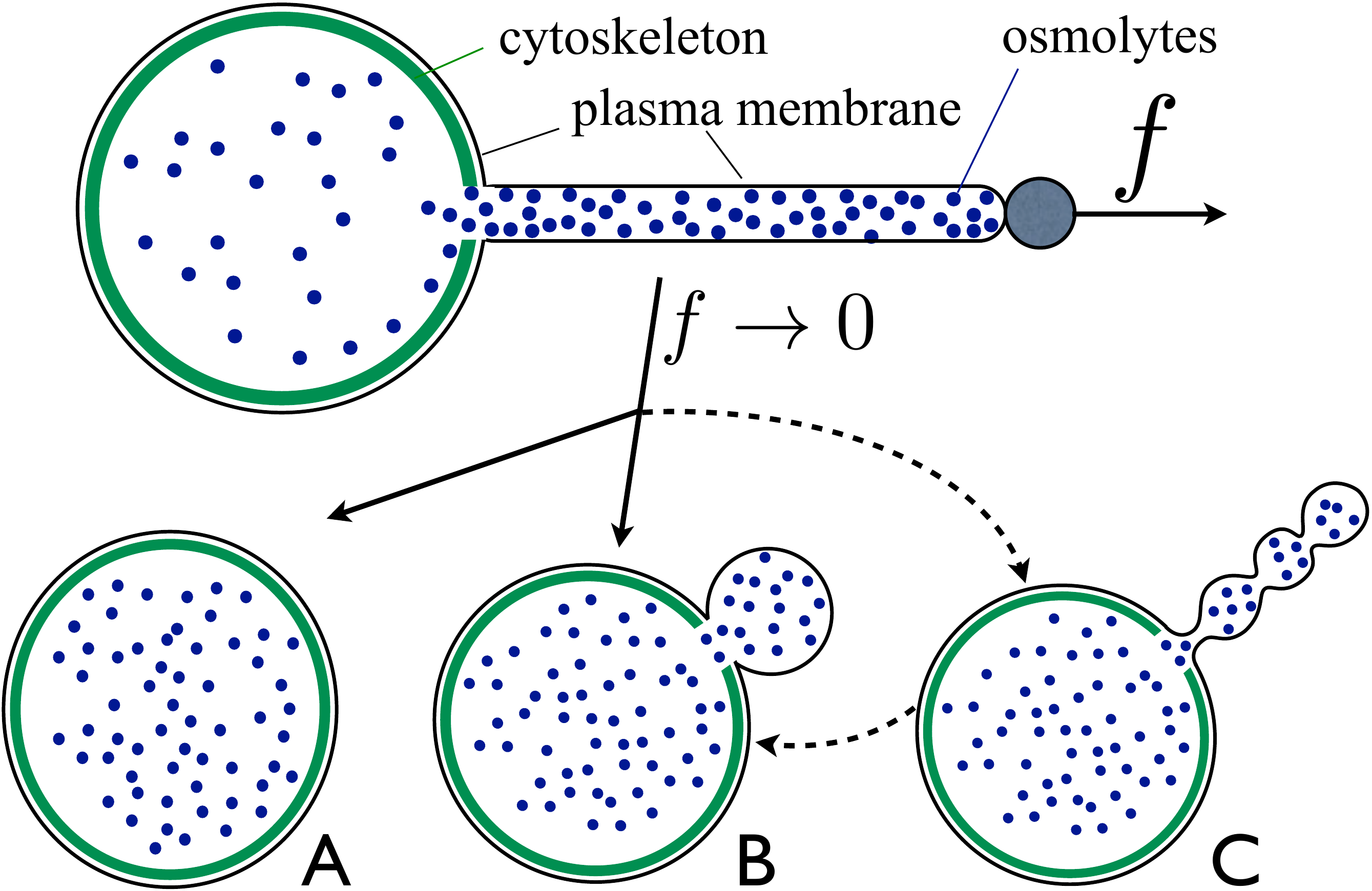}
	\caption{A strongly deformed Red Blood Cell (top) contains a long, cytoskeleton-free membrane tube filled with hemoglobin, created  by an internal force (the polymerization of an hemoglobin fiber) or external force (the action of  optical tweezers). Upon force removal ($f\rightarrow0$) the cell may relax to its initial shape if the tube completely retracts (A), or to a different state where some membrane area remains outside the cell body and forms a spherical bulge  (B). In the latter case, the tube might exhibit pearling (C) during relaxation.} \label{sketch}
\end{figure}

Our goal is to determine the conditions under which a transient mechanical perturbation can give rise to irreversible cellular modifications. We study the model sketched \fig{sketch} of a highly deformed RBC presenting a tubular protrusion and investigate the cell's relaxation after the force that created the protrusion is removed. Such deformation can be generated by the polymerization of long fibers \cite{kuchnir:1997b}, or the application of a localized mechanical force on the cell membrane \cite{hochmuth:1973}.
 We first show that the tight mechanical coupling between the CSK and the cell membrane can generate mechanical frustration and the appearance of several meta-stable cell shapes. We then show that the kinetics of shape relaxation strongly influences the relaxed shape. Finally, we briefly discuss  more complex relaxation routes, including the peculiar pearling of a long and thin protrusion. 

{\bf Metastable shapes of a RBC} The interplay between CSK elasticity and the semi-permeable nature of the PM (permeable to water, but not to Hg or other larger molecules) is known to give rise to a rich phase diagram of equilibrium shapes, which can be explored by varying external parameters such as osmolarity or temperature \cite{lim:2008}. Here, we are not concerned by this kind of equilibrium shape transition, but wish to study the existence of alternative metastable shapes, where some membrane separates from the cell body, or fail to  reincorporate it when extracted. For this purpose,  we assume that the reference state for the cell is a simple sphere of volume ${\cal V}$, and that the total cell volume remains constant during shape transformation (we have check that relaxing this constraint does not alter our results). The elastic energy of the reference state includes contribution for the  elastic energy of the CSK (assumed quadratic: $=K_c({\cal S}-S_0)^2/2S_0$, with a stiffness $K_c$ and a reference area $S_0$, and where ${\cal S}=(36\pi)^{1/3}{\cal V}^{2/3}$ is the CSK area). 
It also includes a contribution for the plasma membrane, which we simply write $=\sigma {\cal S}$, assuming the membrane tension  $\sigma$ to be constant for simplicity. We have checked that including a more accurate membrane elasticity \cite{evans:1990} does not alter our conclusions.

\begin{figure}[t]
\includegraphics[width=8.5cm]{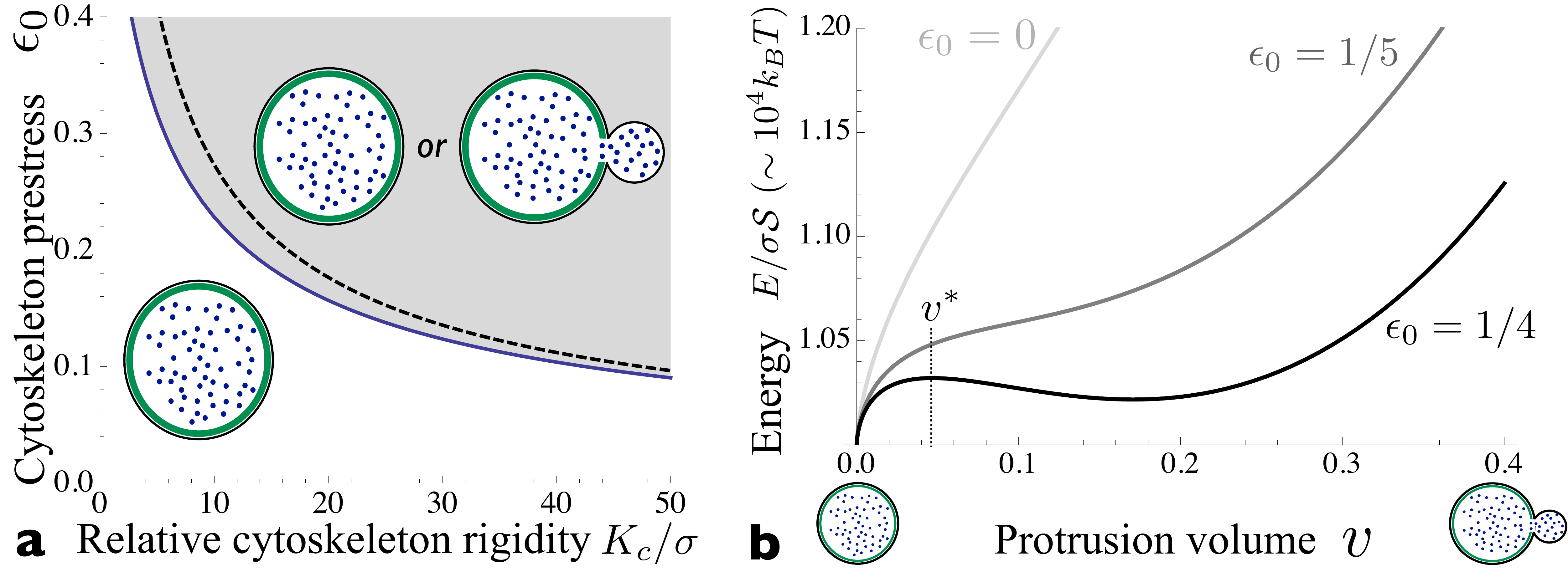}
	\caption{{\bf a)}  Static phase diagram in the parameter space $K_c/\sigma$ (elastic ratio) and $\epsilon_0$ (the CSK prestress), showing the region where a (meta) stable protrusion can exist (shaded grey). The dashed line is the approximation given in the text. {\bf b)} Energy of a RBC with a spherical protrusion as a function of the relative protrusion volume $v$, for different  $\epsilon_0$ ($K_c/\sigma=10$). An energy barrier exist at $v=v^*$ beyond a critical prestress.} \label{static}
\end{figure}

The CSK  is known to contract when the entire RBC membrane is removed \cite{tuvia:1998}, which suggests that it is stretched by its attachment to the cell
membrane (${\cal S}>S_0$). The formation of a membrane bulge detached from the CSK would thus reduce the CSK stretching energy, at the expense of the membrane deformation energy \cite{iglic:1995,sens:2007}. The energy of a protrusion of area $S$ and volume $V$ connected to a spherical body of volume  ${\cal V}-V$  is
\begin{eqnarray}
	E_{\rm el}&=&\sigma\left(S+(36\pi)^{1/3}(\mathcal{V}-V)^{2/3}\right) \nonumber \\
	&+&\frac12 K_c\frac{\left((36\pi)^{1/3}(\mathcal{V}-V)^{2/3}-S_0\right)^2}{S_0}
	\label{Eel}
\end{eqnarray}
The energy of a cell with a spherical protrusion $E_{\rm sph}=E_{\rm el}|_{S=(36\pi V^2)^{1/3}}$, presented \fig{static}, shows the existence of two (meta) stable cell shapes if the CSK is sufficiently stiff or sufficiently stretched by the cell membrane in the reference state. The latter effect is characterized by the CSK prestress parameter: $\epsilon_0\equiv 1-(S_0/{\cal S})^{3/2}$ (the excess cell volume compared to the optimal volume for vanishing prestress). The location of the critical line is approximately: $\epsilon_0=2^{5/4}(\sigma/K_c)^{3/4}-3\sigma/2K_c$. In practice, $\sigma/K_c\simeq 0.1$ ($\sigma\simeq 10^{-6}\unit{J/m^2}$ and $K_c\simeq 10^{-5}\unit{J/m^2}$) \cite{lim:2008}) and extracellular membrane protrusion can be stable  at fairly weak prestress ($\epsilon_0\simeq 25\%$). The shape transition is second order, characterized by a large ($\sim 500\kT$) energy barrier. It does not occur spontaneously near the critical line, but can be triggered, {\em e.g} by an external force. 

{\bf Relaxation kinetics} We assume that a localized force has created a long tubular protrusion of length $L$ and radius $r_i$ (with $L\gg r_i$). The force is switched off at time $t=0$, and the perturbation relaxes. We assume for now that the protrusion shape relaxes smoothly toward  toward a spherical bulge (without pearling), and can be described by two parameters (its volume $V$ and area $S$) continuously evolving from the initial values of a thin tube ($V_i\sim r_i^2L$ and $S_i\sim r_iL$) to those of a spherical protrusion ($V\sim S^{3/2}$). Cytosol volume and membrane area are transferred between the cell body and the protrusion during relaxation, leading to energy dissipation. If dissipation is dominated by  cytosol hydrodynamics, volume exchange is slow and the protrusion area  decreases under almost constant volume to form a small sphere. If dissipation is dominated by membrane flow, the protrusion volume increases under almost constant area to form a large sphere. When the protrusion becomes a sphere, the system evolves in the energy landscape shown \fig{static}, which may exhibits an energy barrier for a protrusion volume $v^*$ ($v=V/{\cal V}$ is the normalized protrusion volume). The initial relaxation dynamics thus determines on which side of the barrier the system  falls, and whether the final state is an intact cell ($v<v^*$), or a cell with a ``permanent'' spherical bleb ($v>v^*$).

In order to study this situation more quantitatively, we write the balance of generalized elastic and dissipative forces using a Lagrangian description \cite{goldstein:2002,pio:vesdyn}. 
\be
	\frac{\partial E}{\partial\{S, V\}}+\frac{\partial \cal P}{\partial\{ \dot S, \dot V\}}=0 
	\label{lag}
\ee
including elastic forces derived from the system's energy $E$ and dissipative forces derived from the energy dissipation (per unit time)  $\cal P$. The energy includes the CSK and membrane tension energy $E_{\rm el}$, \eq{Eel}, and a contribution from the membrane bending rigidity $\kappa$, approximated by the bending energy of a tubular protrusion (the only limit where it is relevant): $E_\kappa\simeq\kappa/8S^3/V^2$.

The energy dissipation functions for membrane and cytosol flows are derived in detail in the Supplementary Information (S.I.). The former is dominated by the membrane flow over the (immobile) cytoskeleton inside the cell body as the protrusion area varies. It  is characterized by a friction parameter $\alpha_{\rm l} \approx[(10^7-10^{11})]\unit{Pa\cdot s/m}$  which dependent logarithmically on the system's size \cite{brochard-wyart:2006}. The latter is dominated by the cytosol flow through the neck connecting the body and the protrusion, assumed to be of small dimension of order the CSK mesh-size $r_n\sim 100\unit{nm}$ \cite{Note3}. It is characterized by a viscosity $\eta\approx \unit[10^{-2}]{Pa /m }$ \cite{evans:1994}:
\begin{align}
	\mathcal P_V= \frac{\eta}{\pi r_n^3} \left(2\dot V -r_n\dot S\right)^2 \quad ,\quad \mathcal P_S= \frac{\alpha_{\rm l}}{4\pi}\dot S ^2. \label{diss}
\end{align} 

Inserting $E=E_{\rm el}+E_\kappa$, \eq{Eel}, and ${\cal P}={\cal P}_V+{\cal P}_S$, \eq{diss}, into \eq{lag} yields two coupled dynamical equations for $S$ and $V$.
Using normalized variables and omitting  various numerical prefactors and less important terms for readability ($\bar x$ is the normalized form of $x$, all details and the complete equations can be found in the S.I.), these equations read:
\begin{eqnarray}
\dot s &=& -\frac{1}{\tau_s}\left(1+\bar\kappa\left(\frac sv\right)^2\right)\label{dyneq}\\
\dot v &=&\frac{1}{\tau_v}\left[\frac{1}{(1-v)^{1/3}}\left(\bar K_c\frac{(1-v)^{2/3}-s_0}{s_0}+1\right)+\bar\kappa\left(\frac sv\right)^3\right]\nonumber
\end{eqnarray}
where $s$ and $v$ are the area and volume of the protrusion (normalized by those of the intact cell) and $s_0=(1-\epsilon_0)^{2/3}$ is the optimal CSK area. The  important elastic parameter in \eq{dyneq} is  the ratio of CSK to membrane elasticity $\bar K_c=K_c/\sigma$ ($\simeq 10$). The bending rigidity $\bar\kappa=\kappa/(\sigma{\cal S})$ only intervenes at the very early stage of tubule relaxation. The important dynamical parameters are the typical exchange times for area ($\tau_s\sim \alpha_{\rm l}{\cal S}/\sigma$) and volume ($\tau_v\sim\eta{\cal S}^2/(\sigma r_n^3)$). The latter depends quite strongly on the size $r_n$ of the neck connecting the protrusion and the cell body. Interpreting \eq{dyneq} is rather straightforward; once the force that caused the protrusion disappears, area is drawn toward the cell body ($\dot s<0$) by membrane tension and to some extent by membrane bending stress, volume is pushed out of the cell ($\dot v>0$) by the CSK (if it is extended - $v<\epsilon_0$), by the membrane tension, and to some extent by the bending stress in the protrusion.

\begin{figure}[t]
		\includegraphics[width=8.7cm]{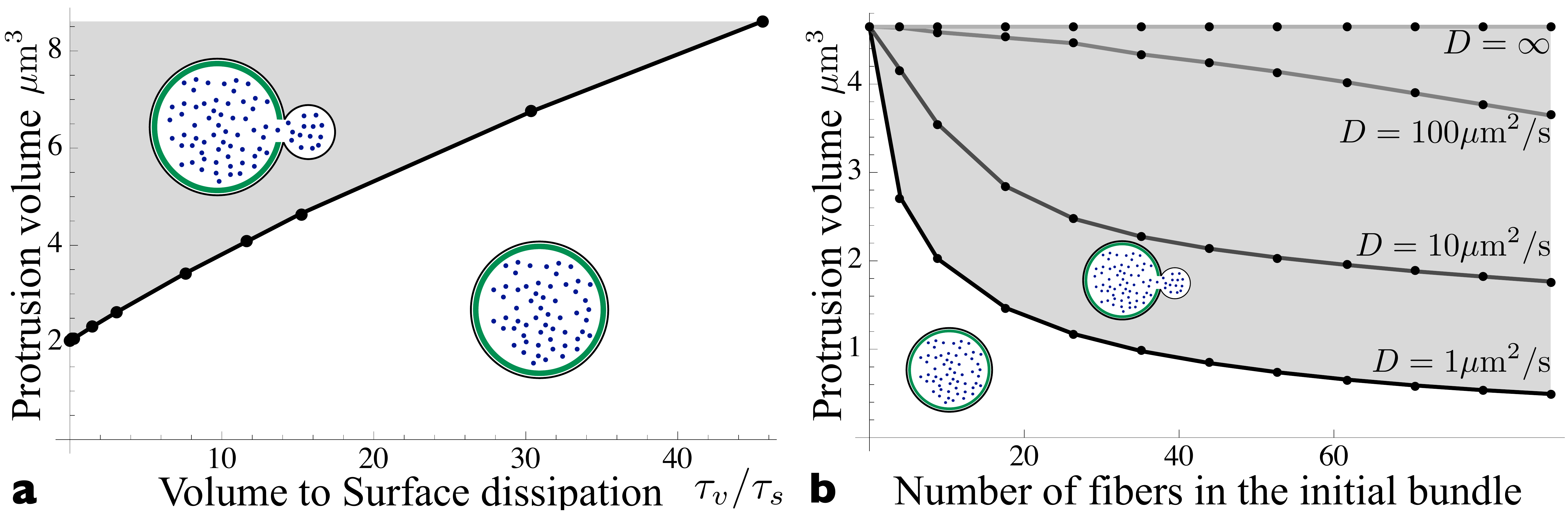}
	\caption{Critical initial protrusion volume beyond which a transient protrusion fails to retract inside the cell body; {\bf a)} for a protrusion created by an external force, as a function of the ratio of the volume to area exchange times (with $K_c/\sigma=10$, $\epsilon_0=1/4$), and {\bf b)} for a protrusion created by a bundle of protein fibers  (proteins of size $4$nm), with $\tau_v/\tau_s=15$), as a function of the number of fibers in the bundle, for different values of the protein diffusion coefficient.} \label{dynamic}
\end{figure}

The dynamical equations \eq{dyneq} are highly non-linear and must be solved numerically. A spherical protrusion is formed once $s(t)=v(t)^{2/3}$ (we call the corresponding volume $v_{\rm x}$), at which point $v$ and $s$ are no longer independent, and the protrusion evolves as a sphere described by a single dynamical equation (given in the S.I). The shape of a cell with a spherical protrusion evolves in the energy landscape of \fig{static}, which possesses two minima and a barrier at $v=v^*$ for large enough prestress. If $v_{\rm x}>v^*$, the deformation is irreversible and a permanent protrusion remains after relaxation.  \fig{dynamic}.a illustrates the impact of the cell dynamics on the critical tether volume beyond which cell shape change is irreversible.
 As discussed earlier, the protrusion is most likely to relax toward a large bulge with $v_{\rm x}>v^*$ if cytoplasm exchange with the cell body is fast compared to membrane exchange. For very fast cytosol exchange ($\tau_v/\tau_s\rightarrow 0$), the protrusion evolves with a constant area, and the irreversible shape change can be triggered by extracting a protrusion of volume $v_i>v^*\simeq 1\%$ of the cell volume. This can be done by extracting a $100$nm-radius membrane tether of length $L\sim100\mu$m  \cite{Note4} (with ${\cal V}=100\mu$m$^3$), a large value, but within experimental range. Slower cytosol to membrane dynamical ratio (increased  $\tau_v/\tau_s$) renders cell deformation more reversible.  For RBC, on expect $\tau_v\sim\tau_s$ and the relaxation time to a spherical protrusion taking up $1\%$ of the total cell area is of order $10$ seconds  (see S.I).



{\bf The case of a depolymerising fiber} In the physiologically relevant case of a protrusion formed by the polymerization of a protein fiber (an Hg fiber in sickle cell anemia), which is then rapidly depolymerize at $t=0$, the Hg concentration and the osmotic pressure are initially be vastly larger in the protrusion than in the cell body. There is a strong driving force to increase  the protrusion volume, and irreversible shape changes are much more likely. Lets consider a protrusion containing $N$ of the total ${\cal N}$ monomers of the cell. Assuming that the density difference between the two compartments vary over the size $r_n$ of the neck connecting the protrusion to the cell body ($r_n\ll L$), Fick's law predicts an equilibration kinetics of the form: $\dot N\simeq Dr_n(\frac{N}{V}-\frac{\cal{N}-N}{{\cal V}-V})$. The normalized kinetic equation for $n\equiv N/{\cal N}$ reads
\be
\dot n=-\frac{1}{\tau_D}\left(\frac{n}{v}-\frac{1-n}{1-v}\right)\quad;\quad \frac{1}{\tau_D}\sim \frac{Dr_n}{\cal V}
\label{Ndyn}
\ee
The system's free energy appearing in \eq{lag} must also include the entropic contribution of both compartments: $E=E_{\rm el}+E_\kappa+E_N$ with (using the ideal gas law)
\be
E_N=\kT {\cal N}\left(n\log \frac{n}{v}+ (1-n) \log \frac{1-n}{1-v}+{\rm cst.}\right)
\label{etot}
\ee
Immediately after fiber depolymerization, the osmotic pressure in the protrusion is very large, leading to a fast inflation of the tether (by volume transfer from the cell body with our assumption of constant total volume ${\cal V}$). This additional force thus increases the likelihood of ending with a permanent protrusion, and the more so if {\em i)} the protrusion volume is large, {\em ii)} the protein density in the protrusion is large, and {\em iii)} protein diffusion through the connecting neck is slow: $\tau_D/\tau_s\gg1$. In sickled RBC, protrusions are formed by Hg fibers  bundled together.
\fig{dynamic}.b shows  the minimal volume a protrusion must have in order not to retract after fiber depolymerization as a function of the number of fibers in the bundle, for different values of the protein diffusion coefficient. 

Note that water can permeate through lipid membranes \cite{olbrich:2000}, and the more so through the membrane of cells, which contain water channels \cite{agre:1991}. Water permeation can be readily included in ${\cal P}_V$ in \eq{diss} (with ${\cal P}_{\rm perm}\sim\alpha_p\dot V^2/S$; $\alpha_{\rm p}$ a permeation coefficient), and does not qualitatively modify our conclusions. It is however likely to have a strong quantitative effect if the  osmotic pressure difference between the protrusion and the extracellular medium is large.   Increase of the protrusion's volume by  osmotic water influx  should strongly promote irreversible shapes change for fiber-induced protrusions or under hypotonic conditions.  

{\bf Pearling of a RBC tether} The previous analysis showed that if a sufficiently long membrane tether is extracted from the cell by a transient force, elastic stress in the CSK can prevent the membrane to reintegrate the cell when the force is switched off, eventually leading to the formation of a ``permanent'' spherical protrusion. The protrusion may however not relax directly to a spherical shape. Experiments using optical tweezers have shown evidence of a pearling instability, during which the tube grows as a necklace of slowly inflating bubbles \cite{hochmuth:1973,Note2}. 

The most favorable case for irreversibility is when relaxation occurs with little area exchange ($S\simeq$ const.). If the areas of both leaflets of the protrusion membrane (the PM is a lipid bilayer) are also constant, any variation of the initial curvature $C_i=1/r_i$ of the protrusion creates differential stretching in the two leaflets, with an energy  $E_{BL}=K_s S/2(hC-hC_i)^2$, where $C$ is the local membrane curvature, $h\ (\simeq 5\unit{nm}$) the membrane thickness and $K_s$ ($\simeq 0.1\unit{J/m^2}$) the bilayer stretching modulus. The differential stretching effect is equivalent to the membrane having a spontaneous curvature $C_i$ which hinders the relaxation toward a spherical protrusion. 

Assuming that the protrusion (initially a cylinder of radius $r_i$, length $L$) relaxes with time as a string of closed-packed bubbles of radius $r(t)=2/C(t)$, the driving force for protrusion growth is $f_V=-\partial E_{BL}/\partial V+P_{\rm eff}$. Here,  $P_{\rm eff}$ includes contribution from the CSK tensile stress, and the osmotic pressure difference with the extracellular medium is water permeation is present. The curvature and volume of a string of spherical bubbles of constant total area are linked by $V\sim L/(C_iC)$. The driving force for protrusion growth thus reads:
\be
f_v=P_{\rm eff}-K_s h^2C^2(C_i-C)
\label{fv}
\ee
If $P_{\rm eff}>K_s h^2C_i^3$, differential stretching is unimportant, and the protrusion grows steadily toward a sphere. If not, the force vanishes for a string of bubble of size of order $r_i$. With $r_i\simeq 100\unit{nm}$, pearling is expected below a pressure of order $1\unit{kPa}$, or an osmotic imbalance of order $1\unit{mMol}$. Protrusions generated by protein fibers or in hypotonic environment should relax directly toward a spherical shape, but a  tether drawn from a cell close to iso-osmolarity should relax through pearling (\fig{sketch}.C), provided it is long enough for the assumption of constant area to be reasonable. This phenomenon is indeed observed experimentally \cite{Note2,hochmuth:1973}. In the absence of any membrane exchange with the cell body, relaxation during pearling is controlled by inter-leaflet lipid exchange (flip-flop) and can be slow. It should be noted that  RBCs, like in most other cells, maintain a preferred lipid asymmetry lipid between the two leaflets by ATP-dependent flipases \cite{lim:2008}. Whether these proteins are mechanosensitive and would naturally actively reduce the bilayer stretching asymmetry is unclear. If this is so, the kinetics of relaxation by pearling  could be strongly ATP-dependent.

\begin{acknowledgments}R. Briehl, P. Nassoy, and M.S. Turner are gratefully acknowledged for stimulating discussion and sharing unpublished data.\end{acknowledgments}

\bibliographystyle{apsrev4-1}

\end{document}